\documentstyle[12pt,aaspp4]{article}
\def\deg{\arcdeg}
\def\etal{{\it et al.}}
\def\eg{{\it e.g.,}}
\def\ie{{\it i.e.,}}

\begin{document}

\title{3C\,48: Stellar Populations and the Kinematics of Stars and Gas
in the Host Galaxy\footnotemark[1]}

\footnotetext[1]{Based on observations made with the NASA/ESA Hubble
Space Telescope, obtained from the data archive at the Space Telescope
Science Institute, which is operated by the Association of Universities
for Research in Astronomy, Inc., under NASA contract NAS 5-26555.}

\author {Gabriela Canalizo\altaffilmark{2} and Alan Stockton\altaffilmark{2}}
\affil{Institute for Astronomy, University of Hawaii, 2680 Woodlawn
 Drive, Honolulu, HI 96822}

\altaffiltext{2}{Visiting Astronomer, W.M. Keck Observatory, jointly operated
by the California Institute of Technology and the University of California.}

\begin{abstract}
We present deep Keck LRIS spectroscopy of the host galaxy of 3C\,48.  
Our observations at various slit positions sample the different 
luminous components near the quasar, including the apparent tidal tail 
to the NW and several strong emission line regions.
  
By fitting Bruzual \& Charlot \markcite{bru96}(1996) population synthesis
models to our spectra, we obtain ages for the most recent major episodes of
star formation in various parts of the host galaxy covered by our slits.
There is vigorous current star formation in regions just NE and SE of the
quasar and post-starburst regions with ages up to $\sim10^8$ years in other
parts of the host galaxy, but most of the NW tidal tail shows no sign
of significant recent star formation.
We use these model fits, together with the kinematics of the stars and
gas, to outline a plausible evolutionary history for the host galaxy, its
recent starburst activity, the triggering of the quasar, and the interaction
of the radio jet with the ambient gas.

There is strong evidence that the 3C\,48 host is an ongoing merger, and that 
it is probably near the peak of its starburst activity.  Nevertheless,
the quasar itself seems to suffer little extinction, perhaps because we
are viewing it along a particularly favorable line-of-sight.

\end{abstract}


\section{Introduction}

Strong interactions and mergers have been implicated in the triggering of
both ultra-luminous infrared galaxies (ULIGs; Sanders \etal\ 
\markcite{san88}1988; 
Sanders \& Mirabel \markcite{san96}1996 and references therein)
and QSOs (\eg\ Stockton \markcite{sto99}1999 and references therein).  
For ULIGs with the highest FIR luminosities 
(${\rm log\ }L_{FIR}>12.0$ in solar units), essentially all appear to be
mergers.  The evidence linking interactions to QSOs, while strong for a
number of specific cases, is generally more circumstantial and 
less compelling than that for ULIGs. This does not necessarily imply lower
interaction rates for QSO host galaxies: it is much
more difficult to observe diagnostic features in QSO hosts because of the
bright nucleus.  Sanders \etal\ \markcite{san}(1988) have suggested that 
there may be an evolutionary path from ULIGs to QSOs.  If so, then signs of the
interaction also would be expected to be less obvious in QSOs because of the
fading and dissipation of tidal features such as tails.

The relation of mergers to intense starbursts (resulting in ULIGs) and to
the feeding of supermassive black holes (resulting in QSO activity) is
undoubtedly complex.  Some mergers of disk galaxies result in only 
comparatively moderate star formation, and most mergers apparently do not
lead to strong nuclear activity.  It is not yet clear whether the
extra ingredients (besides the bare fact of a merger) needed to produce
a sizable starburst are necessarily identical to those required to trigger
strong nuclear activity, but the fact that a fair fraction of ULIGs show
signs of nuclear activity indicates that there is at least a good deal of 
overlap.  In order to explore this connection in more detail, we have 
initiated a program to study stellar populations in the host galaxies of
QSOs having FIR colors tending towards those of ULIGs, and therefore indicating
a substantial contribution from a starburst or recent post-starburst
component.  We have previously fitted population-synthesis models to
high-S/N spectroscopy of the close, 
interacting companion to the QSO PG\,1700+518 (Canalizo \& Stockton 
\markcite{can97}1997;
Stockton, Canalizo, \& Close \markcite{sto98}1998).  Here, we use similar 
techniques to analyze the host galaxy of 3C\,48.

3C\,48 was the first quasar to be discovered (Matthews \etal\ 
\markcite{mat61}1961; Matthews \& Sandage \markcite{mat63}1963). 
As often seems to be the case for the first example of a new class of
astronomical object, it is, in retrospect, far from being typical.
The original identification of 3C\,48 depended in part on
the small size of its radio source, and it remains the only example of
a compact steep-spectrum (CSS) radio source among powerful quasars at redshifts
$<0.5$. The host galaxy of 3C\,48 is unusually large and bright in comparison
with those of other low-redshift quasars (Kristian \markcite{kri73}1973). 
3C\,48 was the first QSO for
which an extended distribution of ionized gas was observed (Wampler \etal\
\markcite{wam75}1975), and it remains one of the most luminous examples of 
extended emission among low-redshift QSOs (Stockton \& MacKenty 
\markcite{sto87}1987).  

Spectroscopy of the host galaxy of 3C\,48 by Boroson \& Oke 
\markcite{bor82}\markcite{bor84}(1982, 1984)
showed strong Balmer absorption lines, demonstrating clearly not only that the 
extended continuum radiation was dominated by stars, but that these stars
were fairly {\it young} ($<1$ Gyr old).  A similar story is told by the 
far-infrared (FIR) colors, which lie between those of most QSOs and those of 
ULIGs, indicating that a significant fraction
of the FIR radiation is likely due to dust in regions of current or recent 
star formation (Neugebauer \etal\ \markcite{neu86}1986; Stockton \& Ridgway 
\markcite{sto91}1991; Stockton \markcite{sto99}1999).  Another connection 
with ULIGs is seen in the apparent tidal tail
extending to the northwest from the 3C\,48 host galaxy and the possibility
that the luminosity peak 1\arcsec\ northeast of the quasar nucleus might be the
nucleus of a merging companion (Stockton \& Ridgway \markcite{sto91}1991).

In this paper, we describe the results of deep spectroscopy of the host
galaxy of 3C\,48, using multiple slit positions to cover most regions of the
galaxy.  After analyzing the stellar velocity field, fitting 
population-synthesis models to the spectra at many discrete points in the
host galaxy, and determining the distribution and velocity structure of the
extended emission-line gas, we discuss the implications of our results in 
terms of a merger scenario.   We assume $H_{0}=75$ km s$^{-1}$ Mpc$^{-1}$ 
and $q_{0}=0.5$ throughout this paper.

\section{Observations and Data Reduction \label{obs}}

Spectroscopic observations of the host galaxy of 3C\,48 were carried out
on 1996 October 13--14 UT and 1996 November 3--4 UT,
with the Low-Resolution Imaging Spectrometer (LRIS; Oke \etal\ 
\markcite{oke95}1995) on the Keck II telescope. 
In Table \ref{journal} we show a complete journal of observations, with
specification of the slit positions.
We used a 600 groove mm$^{-1}$ grating blazed at 5000\,\AA\ for slit positions
A, B, C, and G, yielding a dispersion of 1.28\,\AA\ pixel$^{-1}$.
We used a 900 groove mm$^{-1}$ grating blazed at 5500\,\AA\ for the remaining
slit positions in order to obtain a higher dispersion (0.85\,\AA\ pixel$^{-1}$)
to observe the structure of the emission lines in the spectra.
Total integration times were 5 minutes for slit position D, 40 minutes for 
slit position F, and 60 minutes for all the other slit positions.

The spectra were reduced with IRAF, using standard reduction procedures.
After dark, bias and flat field correction, the images were
background subtracted to remove sky lines.   Wavelength calibrations
were made using the least-mean-squares fit of cubic spline segments
to identified lines in a comparison spectrum.  
The spectra were flux calibrated using spectrophotometric standards from
Massey \etal\ \markcite{mas88}(1988) observed with the slit at the 
parallactic angle.  
The distortions in the 
spatial coordinate were removed with the APEXTRACT routines.
The spectra were not corrected for discrete atmospheric absorption
features; thus they show the atmospheric B-band redwards of the 
[\ion{O}{3}] $\lambda$\,5007 emission feature.

Spectra from slit position A suffered from strong light scattering from the 
quasar.  
In addition, one of the three 20 minute exposures taken at this position
was unusable because of problems with the Keck II mirror control system. 
 
Spectra from slit positions A, B, C and G were subdivided into eight regions
each (see Fig.~\ref{master}$a$) for the analysis of stellar populations and
kinematics of the host galaxy.   The size of each of these regions was chosen 
so that each spectrum would have sufficient signal-to-noise ratio for modeling
while still providing good spatial resolution.

The spectra of regions close to the quasar nucleus were contaminated by 
scattered quasar light.   The scattered light was removed by subtracting 
from each region a version of the quasar nuclear spectrum,
scaled to match the broad-line flux.   The percentage in flux (at 4500 \AA, 
rest wavelength) subtracted
in each case is listed in column (6) of Table \ref{regions}.
This quantity depends both on the intensity of the quasar profile wings and 
the surface brightness of the host galaxy in each particular region.

WFPC2 images of 3C\,48 were obtained from the {\it Hubble Space Telescope} 
({\it HST}) data archive.
The images used in this analysis included two 1400 s exposures in the F555W
filter and two 1700 s exposures in the F814W filter.  In each case, 3C\,48
was centered on the PC1 detector.  We also used two 1100\,s and one
1300 s exposures in the linear ramp filter FR680N, centered on redshifted
[\ion{O}{3}] $\lambda$\,5007.  3C\,48 fell near the center
of the WFC2 chip.  All {\it HST} images were reduced as described
in Canalizo, Stockton \& Roth \markcite{can98}1998.  We subtracted a 
scaled stellar profile
at the position of the quasar, using Tiny Tim (Krist \markcite{kri93}1993)
models of the {\it HST}/WFPC2 PSF.

We also use three ground-based images of 3C\,48 obtained with the University 
of Hawaii 2.2 m telescope.  The first of these is a sum of three 2700\,s 
exposures,
taken on 1991 September 16 with a Tektronix 1024$\times$1024 CCD through a 
30\,\AA\ bandpass filter centered on redshifted [\ion{O}{3}] $\lambda$\,5007, 
with an image scale of 0\farcs22 pixel$^{-1}$.
The second is a sum of 30 300\,s exposures, taken on UT 1993 September 18 
with a Tektronix $2048\times2048$ CCD through a filter centered at 6120\,\AA\ 
and having a 960\,\AA\ FWHM, covering the relatively line-free region of the
rest-frame spectrum from 4120 to 4820\,\AA\ in the rest frame of 3C\,48,
also at a scale of 0\farcs22 pixel$^{-1}$.  We shall refer to this filter
as ``$R'$.''  Finally, we obtained short wavelength images of 3C\,48 on 
UT 1998 September 18 through a $U'$ filter
(centered at 3410\,\AA\ with a 320\,\AA\ FWHM). 
Nine 1200\,s exposures were taken using an Orbit CCD, with an image scale 
of 0\farcs138 pixel$^{-1}$. 
All ground-based images were reduced with IRAF using standard 
image reduction procedures.  The $U'$ and $R'$ have been processed with
CPLUCY (Hook \markcite{hoo98}1998), an improved version of the PLUCY task 
(Hook \etal\ \markcite{hoo94}1994)
available in the STSDAS CONTRIB package.  Both tasks carry out a two-channel
deconvolution of a field, one channel comprising designated point sources, 
which are deconvolved using the standard Richardson-Lucy algorithm, and the 
second channel
including everything else, for which the Richardson-Lucy deconvolution is
constrained by an entropy term to ensure smoothness.  One of the virtues of
this procedure is that, with a proper choice of parameters, it is possible to
remove the stellar profile from a non-zero background without the ringing
problem inherent in the application of the standard Richardson-Lucy procedure.

Both the ground-based images and the {\it HST} images are shown in 
Fig.~\ref{imagemontage}, to which we will refer as needed during our 
discussion.
 
\section{Results}
 
\subsection{Kinematics of the Stellar Component in the Host Galaxy 
\label{kinematics}}
 
Redshifts determined from stellar absorption features give us information 
about the kinematics of the stars in the host galaxy.
We have obtained measurements from 32 regions covered by 4 slit positions
(A, B, C and G; see Table 1).   Some of these regions overlap, 
providing a consistency check.  Column (2) of Table \ref{regions}
lists the relative velocity of each region with respect to $z = 0.3700$, 
which is close to the average of the
stellar redshifts measured in the inner regions of the host.

Measuring redshifts from stellar absorption features presents two main
problems: the absorption features may be contaminated by emission lines 
coming from the extended ionized gas, and some absorption features may 
be interstellar in nature. 

Contamination of absorption features by emission can yield a higher or lower
redshift depending on whether the ionized gas has a lower or higher redshift 
than the
stars.   In order to check for contamination, we examined the Balmer lines. 
Because the Balmer decrement is steeper in emission than in absorption, 
those lines that are higher in the series (\eg\ H8, H9, H10) show less 
contamination by emission.
As a result, in the regions where there is extended emission, 
the redshift measured from the lower lines in the Balmer series was
different from that measured from the higher-series 
lines.
This effect is quite significant in some cases, and it correlates 
with the strength of the emission observed in our [\ion{O}{3}] image
(Fig.~\ref{imagemontage}$f$).
Where there was evidence for strong emission contamination,
we measured the redshift of the stellar component only from the higher 
Balmer lines (usually H 8 and above) and the \ion{Ca}{2} K line.
We did not measure redshifts from \ion{Ca}{2} H when a young stellar
component was present as this line was blended with H\,$\epsilon$.

Redshifts measured from \ion{Ca}{2} lines, however, must be treated with
caution, as these lines occur often in the interstellar medium.
From our modeling of stellar populations (see \S~\ref{pops})  we notice
that \ion{Ca}{2} is sometimes stronger in the observed spectrum
than in an otherwise good fitting model
(\eg\ at B4 in Fig.~\ref{typical}), and this is indicative of intervening
interstellar gas, which may have a different velocity from that of
the stellar component.   However, for those spectra with a young stellar 
component, the redshift we measured from \ion{Ca}{2} K was always consistent 
with that measured 
from the higher members of the Balmer series within measurement errors.
In the cases where there was no young stellar component, we checked
for consistency with other old population features, such as the  
the \ion{Mg}{1}$b$ feature.

Figure \ref{master}$b$ shows a map of the radial velocities in the host galaxy.
The north-west tidal tail-like extension of the host galaxy is blueshifted 
with respect to the main body of the host galaxy by as much as 
300 km s$^{-1}$.   
The west end of the tail has smaller approaching velocities 
($> -$200 km s$^{-1}$ ), which increase to a maximum of 
$\sim -$300 km s$^{-1}$, and then decrease again ($\sim -$100 km s$^{-1}$)
as we approach the main body of the host.
We find relatively small variations in velocity in the central region of the
host galaxy.  However, the south-east part of the galaxy appears to have
receding velocities as high as $+$180 km s$^{-1}$.   
This may indicate that the system is rotating around an axis oriented
roughly northeast---southwest.

\subsection{Stellar Populations \label{pops}}

\subsubsection{Modeling the Spectra}

The high signal-to-noise ratio of the Keck LRIS spectra allows us to attempt 
modeling different regions of the host galaxy, with sizes ranging down to
approximately one square arcsecond in the higher-surface-brightness regions.
We have modeled 32 individual regions
(see Fig.~\ref{master}$b$) of the host galaxy using the 
Bruzual \& Charlot \markcite{bru96}1996 synthesis models.
% (we have omitted regions A3 and
%A4 because we were unable to obtain a satisfactory correction for the
%scattered light from the quasar at these locations).

Typical spectra of the host galaxy are shown in Fig.~\ref{typical}.
The spectra generally show Balmer absorption lines from a relatively young 
stellar population as well as evidence for an older population, such as the 
Mg\,I{\it b} feature and the \ion{Ca}{2}\,K line.   The simultaneous presence
of both kinds of features indicates that we 
can do some level of decomposition of the two components.

As we shall discuss in more detail in \S~\ref{merge}, 
the morphology of the host galaxy indicates that some form
of strong interaction has occurred.  We assume that this strong tidal 
interaction has induced one or more major starburst episodes in the host 
galaxy.  Thus, to first order, the observed spectrum consists of an 
underlying old stellar population (stars in galaxy before interaction) plus 
a recent starburst (stars formed as a result of interaction).

Our early experiments in modeling the old population in similar situations
showed that, as expected, it is not strongly constrained by the observations.  
On the other hand,
we have found that the precise choice of the underlying old stellar population 
makes very little difference in the modeling of the superposed starburst 
(Stockton, Canalizo \& Close \markcite{sto98}1998).  
Accordingly, we have made what we take to be physically reasonable assumptions
regarding the old population and have used this model for all of our analyses.
In our early modeling of the central regions
of the host galaxy the old stellar component was well fit by a 10 
Gyr old population with an exponentially decreasing star formation rate with an
e-folding time of 5 Gyr.   Later, we were able to extract a purely old 
population (not contaminated by younger stars) in the tail, far from the
nucleus (see C8 in Figs.~\ref{master}$a$ and \ref{typical}).   The old 
population model we had previously chosen gives a very reasonable fit,
confirming that we had chosen a reasonable model for the old stellar 
population.

We assume that the same old stellar population is present 
everywhere in the host galaxy.  To this population we add isochrone
synthesis models (Bruzual \& Charlot \markcite{bru96}1996) of different ages.
The assumption of a 
population with no age dispersion can be justified if the period during
which the star formation rate was greatly enhanced is short compared to
the age of the population.    Indeed, observational evidence
indicates that the duration of starbursts in individual star forming clouds
can be very short, often 
$< 1$ Myr (\eg\ Heckman \etal\ \markcite{hec97}1997; Leitherer \etal\ 
\markcite{lei96}1996).
We perform a $\chi^2$ fit to the data to determine the contribution of 
each component and the age of the most recent starburst.   

Since the size of each region we are modeling is at least 
$1\times1\arcsec$ ($\sim 4\times4$ kpc), 
the observed spectrum is likely to be the integrated spectrum
of several starbursts of different ages.  The starburst age we derive for
the observed spectrum will be weighted towards the younger starbursts
because such starbursts will have a greater flux contribution.
On the other hand, this weighted-average age will likely be at least somewhat
greater than the age of the youngest starburst in the observed region.
Thus our ages can be taken as upper limits to the age of the most recent
major episodes of star formation along the line of sight.

While the starburst ages that we list on Table 2 represent the best
fit to the observed spectrum, there is a range of ages in each case that
produces a reasonable fit.   Figure \ref{agerr} shows the spectrum from region
B8 with three different solutions superposed:  The youngest that still looks
reasonable (top), the best fit (middle), and the oldest that still looks
reasonable (bottom).  In each case, the relative contribution of the old 
population was increased or decreased to give the best fit.
The difference in $\chi^2$ statistic between the middle and top or middle and 
bottom fits is $\sim$15\%.
We take these limits to be an estimate of the error in the determination
of age so that, for this particular case, the starburst age would
be $114\ (+67, -42)$ Myr.  In general,  a reasonable estimate 
of the error for the starburst age determined in each case is $\pm50\%$.

As mentioned in \S~\ref{kinematics}, absorption lines in many regions
are contaminated by emission lines coming from the extended ionized gas.
In modeling these regions, we excluded those Balmer lines that were
obviously contaminated by emission from the $\chi^2$ fits.

Notice that in some of the spectra (\eg\ G2 in Fig. \ref{typical}), the 
observed data are slightly depressed with respect to the models in the 
region between 4600 \AA\ and 4800 \AA .   We previously found this problem 
in the spectrum of the companion to PG\,1700+518 (Canalizo \& Stockton 
\markcite{can97}1997) and attributed it there to an artifact of the 
subtraction of the QSO scattered light.  However, some of the spectra of 
3C\,48 show this problem even when no QSO light was subtracted.  
We find no evidence for a correlation of this effect with the age of the 
starburst.
Fritze-von Alvensleben \& Gerhard \markcite{alv94}(1994) seem to have run 
into the same problem when they use their own models to fit a spectrum of 
NGC\,7252 (see their Fig.~1).  This problem is also evident in the spectrum 
of G\,515 in Liu \& Green \markcite{liu96}1996.   Neither of these groups 
discusses the discrepancy.  

\subsubsection{Mapping Stellar Populations}

Figure~\ref{master}$c$ shows a map of the host galaxy with the starburst
ages we determine from spectra.
Figure~\ref{typical} 
shows a characteristic spectrum of each of the regions described below.  
 In column (3) of Table~\ref{regions} we list the ages of 
the young stellar component for each region analyzed.  In columns (4)
and (5) we give, respectively, the mass and light (at 4500\AA, rest wavelength)
contribution of the starburst to the observed spectrum.

We start with the long extension to the north-west of the quasar which is
very likely a tidal tail, as we shall see.
As we mentioned in the previous section, the stellar populations in most of 
this tail appear to have no younger component.   Panel C8 of 
Fig.~\ref{typical} shows one such spectrum
with a $\chi^2$ fit to a 10 Gyr-old population with an exponentially 
decreasing star formation rate with an e-folding time of 5 Gyr.  

Even though
most of the extension is made up by old populations, there is one
small starburst region covered by our slit positions G and A.
This starburst has an age $\leq$ 33 Myr and is about 4\arcsec\
west and 8\arcsec\ north of the quasar nucleus, where A7 and G8 intersect
(see Fig.~\ref{master}$a$,$c$).   A faint clump at the same position
is clearly seen on our $U'$ ground-based image (Fig.~\ref{imagemontage}$d$).

The $U'$ image also shows two additional larger bright regions north of 
the quasar nucleus.  The eastern region corresponds to a strong emission region
clearly seen in the [\ion{O}{3}] image shown in Fig.~\ref{imagemontage}$f$.   
The western 
region (covered by C6 and adjacent regions in Fig. \ref{master}), on the 
other hand, is much 
fainter in both the narrow-band and broad-band images than in the 
$U'$ image.   This region contains starbursts of ages $\sim 9$ Myr, and
the F555W {\it HST} image suggests that it is formed of several smaller
clumps.

We also find starburst ages $\sim 9$ Myr in the regions directly north 
(C3, C4) and southwest (G3) of the quasar.  These are seen in the $U'$ image 
as relatively high surface brightness areas.    The area corresponding to 
G3 has a knotty structure, which is not obvious in the longer wavelength 
images.

We have tried taking the ratio of our $U'$ image to our $R'$ image 
(Fig.~\ref{imagemontage}$e$).  Both of these images
should be strongly dominated by continuum radiation; however, the
physical interpretation of this ratio is not straightforward 
for two reasons:  (1) 
while the $U'$ image should relatively emphasize regions that have young stars,
it may also be enhanced by strong nebular thermal emission from emission
regions, and (2) as our modeling of the stellar populations shows, the ratio
of young and old stellar populations in the $R'$ image is highly variable.  
Nevertheless, a striking feature of the ratio image is the apparent ridge
of strong UV radiation along the leading edge of the NW tail.  This is 
apparently a complex of emission regions and young stars, including both
the brightest emission region seen in the [\ion{O}{3}] images and the A7/G8
star-forming region mentioned above.

On the west side of the nucleus (along Slit G) we find that there is no
recent star formation in the northern half of the slit other than the 
star forming clumps mentioned above.  
The southern part of the host galaxy seems to be dominated by relatively
older starbursts (generally $\gtrsim$100 Myr) with a range of ages between
16 and 114 Myr.

Finally, the youngest populations we observe in the host galaxy are found 
in the regions coincident with the two brightest features in the
$U'$ image (see inset in Fig.~\ref{imagemontage}$d$): (1) a region 
$\sim2\arcsec$ SE of the quasar, covered by A3 
and B5, and (2) a region $\sim1\arcsec$ NE of the quasar, surrounded by
B3 and C2, which will be discussed in more detail in \S \ref{jet}.
We find starburst ages as young as 4 Myr in these regions.   
Such young ages are comparable to starburst 
timescales and it is virtually impossible to distinguish between 
continuous star formation and instantaneous bursts, so the young 
ages we derive simply indicate that there is ongoing star formation.

\subsection{Emission}

\subsubsection{Emission in the Quasar Nucleus}

Thuan, Oke, \& Bergeron\markcite{thu79} (1979) noticed that the broad
permitted emission lines in the spectrum of 3C\,48 had a systematically
higher redshift, by about 600 km s$^{-1}$, than did the forbidden lines.
Boroson \& Oke\markcite{bor82} (1982) confirmed this qualitative result 
but found a lower velocity difference of $330\pm78$ km s$^{-1}$.
Gelderman \& Whittle\markcite{gel94} (1994) mentioned that the redshift
of the forbidden lines is difficult to measure because they ``have
approximately flat tops.''

Our spectrum of the quasar nucleus in the region around H$\beta$ and
[\ion{O}{3}] $\lambda\lambda4959$,5007 is shown Fig.~\ref{hb_nuc_emiss}.
The [\ion{O}{3}] lines clearly have double peaks.  By deconvolving the lines
into best-fit Gaussian components, we find that the narrower, higher velocity
component has a mean redshift $z=0.36942$, with a Gaussian width
$\sigma=400$ km s$^{-1}$, while the broader, lower-velocity component has
a mean redshift $z=0.36681$, with $\sigma=700$ km s$^{-1}$.  These
redshifts correspond to a velocity difference of $563\pm40$ km s$^{-1}$. 
The redshift
of the broad H$\beta$ line is $z=0.36935$, in good agreement with the 
higher-redshift [\ion{O}{3}] line.  Thus the anomalous feature is the broad, 
lower-velocity [\ion{O}{3}] line.

What is the nature of this blue-shifted component?  First, it is highly
luminous: the luminosity in the [\ion{O}{3}] $\lambda5007$ line alone is
$\sim5\times10^{43}$ erg s$^{-1}$, over 3 times the luminosity of the 
same line in the ``standard'' narrow-line
region.  Second, it does show some spatially-resolved velocity structure,
though only over a region of about 0\farcs5.  We first noticed this
velocity gradient in our reduced long-slit spectrum, but it is very difficult
to display because of the strong variation of the underlying quasar continuum
in the direction perpendicular to the slit.  Figure \ref{oiiidecon} shows the
result of a PLUCY (Hook \etal\ \markcite{hoo94}1994) deconvolution of the 
2-dimensional
spectrum, followed by a reweighting of the image perpendicular to the slit
to even out the dynamic range in the resolved emission.  The steep velocity
gradient is clearly visible in the blueshifted component of both of the
[\ion{O}{3}] lines.

As we were completing this paper, an important new study by Chatzichristou,
Vanderriest, \& Jaffe \markcite{cha99}(1999) appeared, in which the 
spatial and velocity structure of the 
emission lines in the inner region of 3C\,48 were mapped with a fiber 
integral-field spectroscopic unit.  They also
observed the double-peaked narrow-line emission in the quasar spectrum,
and their decomposition of the profile gave a radial velocity difference
of $586\pm15$ km s$^{-1}$, with which our value of $563\pm40$ km s$^{-1}$ 
is in good
agreement.  Our results are in general agreement on most other points for
which the two datasets overlap, except that Chatzichristou \etal\ find
the luminous broad blueshifted component 
to have no resolved velocity structure, whereas we find it to have a strong 
velocity gradient over the central 0\farcs5 region.

\subsubsection{Extended Emission}

We have a total of 6 slit positions useful for studying the extended 
emission, which are shown in Fig.~\ref{emslits}, superposed on our
narrow-band [\ion{O}{3}] image.  Images of the two-dimensional spectra
are shown in Fig.~\ref{specimage}.  We have fitted Gaussian profiles to
the [\ion{O}{3}] $\lambda5007$ lines (and to the [\ion{O}{2}] 
$\lambda\lambda3726$,3729 lines in our higher-dispersion spectra from slits
E and F) in order to measure the velocity field of the extended emission.
At some positions, we had to use up to 3 components in order to get a
satisfactory fit to the observed profile.  The results are shown in
Fig.~\ref{emvel}, where the velocities, fluxes, and widths of the lines
are indicated at each position.  We also plot the velocities of the stellar
component for slit positions A, B, C, and G.  
While there is generally at least rough agreement between the velocities
of the stars and at least one component of the emission-line gas, there
are significant differences in detail.   Little, if any, of the emission
can be due to an undisturbed {\it in situ} interstellar medium.
The total velocity range
of the emission is over 1200 km s$^{-1}$, and the FWHM of the emission
reaches up to 2000 km s$^{-1}$.

The morphology of the extended emission, as seen in our deep ground-based
[\ion{O}{3}] $\lambda5007$ image (Fig.~\ref{imagemontage}$f$) and the
{\it HST} WFPC2 linear ramp filter image in the same emission line 
(Fig.~\ref{imagemontage}$i$),
consists of several discrete emission regions having characteristic sizes
ranging from a few hundred pc to $\sim1$ kpc, connected by a web of 
lower-surface-brightness emission.  Some of the emission regions lie
outside the host galaxy, as defined by the deep continuum images.
The {\it HST} image shows an apparent bright object about 0\farcs8 S of the 
quasar, which has no counterpart on the {\it HST} continuum images.
Kirhakos \etal\ \markcite{kir99}(1999) have referred to this object as
an \ion{H}{2} region.  Its absence on the {\it HST} continuum images would
indicate an [\ion{O}{3}] $\lambda5007$ equivalent width
$\gtrsim4000$ \AA.  An equivalent width this large is not 
impossible: if the intensity ratio of [\ion{O}{3}] $\lambda5007$ to H$\beta$ 
is $\sim10$, as is common in extended emission regions around QSOs, the
equivalent width of H$\beta$ would be about a factor of 3 below that which
would be expected from the nebular thermal continuum alone for a temperature
of 15000 K.  
However, we estimate that the object's absence on the {\it HST} PC F555W image
requires that the flux of [\ion{O}{2}] $\lambda3727$ be $<4$\%
that of [\ion{O}{3}] $\lambda5007$, rather than the more typical $\sim20$\%.
%However, the object's absence on the {\it HST} PC F555W image
%also sets strong constraints on the flux in [\ion{O}{2}] $\lambda3727$:
%we estimate that it would have to be $<4$\% 
%that of [\ion{O}{3}] $\lambda5007$, rather than the more typical $\sim20$\%.
Finally, we have carried out
a careful deconvolution of our ground-based [\ion{O}{3}] image, which
has an original FWHM of 0\farcs74 for stellar profiles.  On our deconvolved
image, we could easily have seen any object as bright as the emission region
$\sim2\farcs5$ east of the quasar, which appears almost 10 times fainter than
the apparent object 0\farcs8 south of the quasar on the {\it HST} [\ion{O}{3}]
image, but there is no feature present at this location on our deconvolved
image.  While the WFPC2 Instrument Handbook does not mention the presence
of nearly in-focus ghost images, these doubts about the reality of the feature 
prompted us to check with Matthew McMaster, a data analyst at STScI specializing
in WFPC2 instrument anomalies and reduction of linear ramp filter data.
He examined a calibration image of a star taken with the same filter and
centered on nearly the same wavelength, and he found a similar feature at
exactly the same position in relation to the star, confirming that the apparent
object seen near 3C\,48 is an artifact.

The resolution of the [\ion{O}{2}] $\lambda\lambda3726$,3729 doublet in
the brighter regions of the extended emission is of special interest,
since electron densities can be determined from the intensity ratio
(\eg\ Osterbrock \markcite{ost89}1989).  For the bright pair of regions 
centered 4\arcsec\
N of the quasar (see Fig.~\ref{imagemontage}$i$), we find a ratio
$I(3729)/I(3726)=1.25\pm0.05$, corresponding to an electron density of
150 $e^-$ cm$^{-3}$, assuming an electron temperature $T_e=10^4$ K.
This density is sufficiently high that it is unlikely that the emitting gas, at
its location some 20 kpc projected distance from the quasar, is in
hydrostatic pressure equilibrium with the surrounding medium.  Instead, it is
probably either confined gravitationally (if, for example, the gas is in
a dark-matter dominated dwarf galaxy) or compressed by shocks due to 
collisions of gas clouds during the interaction.  This latter process
seems to be the more likely one, at least for the string of emission and
star-forming regions along the leading edge of the NW tail of 3C\,48.

\section{Discussion} \label{discus}
\subsection{Evidence for a Merger} \label{merge}

Our modeling of the stellar populations in the faint tail-like structure 
extending to the NW of the quasar indicates that this extension is made up 
mostly of old stars.   
The predominance of old stellar populations in a feature that clearly
has a dynamical timescale much shorter than the age of the stars that 
comprise it points strongly to a tidal origin.
(The possibility that the apparent old population is the result of a 
truncated IMF rather than actual old stars can be discounted since we observe
unambiguously high mass stars in some clumps within the tail and elsewhere
in the main body of the host.)
There can be little doubt that the feature is indeed a tidal tail.
If this tail has an inclination angle $i$ between 30\deg and 60\deg
with respect to the plane of the sky, 
the dynamical age of the feature is between 100 and 300 Myr.
Since the oldest starburst ages we find in the host galaxy are 114 Myr, 
the dynamical age of the tail indicates that the starbursts were induced 
after the initial encounter of the interacting galaxies.

If the host galaxy of 3C\,48 has undergone strong gravitational interaction 
with a second object, and both objects originally possessed cold stellar
disks, one might expect to observe also counter-tidal 
features such as a second tidal tail.
Boyce, Disney \& Bleaken \markcite{boy99}(1999) published the archival 
{\it HST} F555W image
of 3C\,48 and identified a second tidal tail ``stretching 8\farcs 5 to the
south-east of the quasar''.  We see a hint of a ``tail'' south of 
the nucleus and extending towards the south-east in both the {\it HST} images
and our $R'$ image, but it is only about $\sim 6\arcsec$ long.
The object that Boyce \etal\ seem to consider the SE end of the ``tail'' is 
a background galaxy at
redshift $z=0.8112$ (object 4 in Table \ref{seren}).  When this object
is removed, the case for this tidal tail becomes considerably less compelling. 

Another possible second tidal tail is the feature starting
SE of the nucleus, but arching from the SE towards the SW.
The broad-band optical ground-based images show a 
relatively bright feature extending $\sim 3\arcsec$ SE of the nucleus 
and the {\it HST} images show peculiar structure around the same area.
This could be the tidal tail with clumps of star formation (see below)
of the merging companion.  Typically, double tails in merging systems
appear to curve in the sense that they are roughly rotationally 
symmetric.
If this SE extension is a tidal tail, the two tails would, instead,
present more of a ``gull-wing'' appearance.  While this is not  
the usual case, a combination of disks inclined to the plane
of the mutual orbit and projection effects can lead to precisely such a
configuration.   The best-known example is ``The Antennae'' (NGC 4038/9;
Whitmore \& Schweitzer \markcite{whi95}1995 and references therein) whose 
configuration has been reproduced in numerical simulations 
(Barnes \markcite{bar88}1988; Toomre \& Toomre \markcite{too72}1972).

Our kinematic results clearly indicate that the NW tidal tail
has approaching velocities with respect to the main body of the galaxy.   
On the other hand, we find some evidence that the south extension(s) has 
receding velocities.  Whether the $\sim3\arcsec$ SE extension is a 
tidal tail and the system has an Antennae-like configuration, or the longer
extension mentioned by Boyce \etal\ \markcite{boy99}(1999) is the tidal tail
forming a more typical merging configuration, the observed velocities are 
qualitatively in agreement with what we would expect from theoretical models 
(Toomre \& Toomre \markcite{too72}1972).
However, there is much uncertainty in the interpretation of these southern 
features, and at this stage we cannot conclusively decide which, if any, of
these features is a real tidal tail.

Our spectroscopy confirms that the clumps in the NW tidal tail observed on
{\it HST} images (Kirhakos \etal\ \markcite{kir99}1999) are regions of 
star formation, with ages $\lesssim33$ Myr. 
Similar clumps have been observed in a number of 
strongly interacting or merging systems such as  NGC\,3628 (Chromey \etal\ 
\markcite{cho98}1998), IRAS\,19254$-$7245 (Mirabel, Lutz \& Maza 
\markcite{mir91}1991), Arp\,105 (Duc \& Mirabel \markcite{duc94}1994) and 
NGC\,7252 (Schweizer \& Seitzer \markcite{sch98}1998). 
Small star forming clumps in the tidal tails of merging systems are 
also predicted by numerical simulations (\eg\ Elmegreen, Kaufman \& Thomasson
\markcite{elm93}1993; Mihos \& Hernquist \markcite{mih96}1996). 

The observational evidence seems to indicate that 3C\,48 is a major merger 
according to the definition of Mihos \& Hernquist 
\markcite{mih94}\markcite{mih96}(1994,1996), that is, 
a merger of two gas-rich galaxies of roughly comparable size.
In addition, at least one of the galaxies probably had a 
significant bulge component.  In the absence of a bulge, merging galaxies 
are predicted to undergo early dissipation, producing central 
starbursts in the disks shortly after the initial encounter, when the 
galaxies are still widely separated, and leaving little gas to form stars 
at the time of the merger (Mihos \& Hernquist \markcite{mih96}1996).  

In the case of 3C\,48, however, we observe that there are strong starbursts
still going on while the galaxies are close to the final stages of merging.
The detection of large amounts of molecular gas in 3C\,48 (Scoville \etal\
\markcite{sco93}1993; Wink \etal\ \markcite{win97}1997), along with the 
very young ages for starbursts that
we observe ($\sim 4$ Myr), indicates that high rates of star formation
are still present in the inner regions.
Merging disk/bulge/halo galaxies have much weaker initial encounter 
starbursts, so that most of the gas is conserved until the final merger
and the rapid collapse of this gas drives a tremendous starburst
in the center of the merging 
system.  

The importance of the presence of a massive bulge in preserving gas for
star formation at the time of the final merger depends on rather uncertain
assumptions regarding the star-formation law.  However, there is little
disagreement that an interaction involving a disk that is marginally
stable against bar formation will lead to enhanced star formation at the
time of the initial passage.  The fact that {\it all} of the starbursts we 
observe were produced after the tidal tail was first launched strongly 
suggests that the presence of a bulge was important in the case of 3C\,48.

\subsection{The Effect of the Radio Jet \label{jet}}

As mentioned in \S1, 3C\,48 is a compact steep-spectrum radio source.  The 
bright radio structure
extends $\sim0\farcs5$ and takes the form of a one-sided jet extending
approximately to the N, but with considerable distortion and local
irregularities (Simon \etal\ \markcite{sim90}1990; Wilkinson \etal\
\markcite{wil91}1991).  Lower-surface-brightness emission
extends in a fan to the E side of this jet and continues to the NE out to
a distance of $\sim1\arcsec$ from the nucleus.  The structure of this
radio emission indicates interaction with a dense gaseous medium
(Wilkinson \etal\ \markcite{wil91}1991).  Is there any evidence in the
optical observations for such an interaction?

We see spatially resolved velocity structure in the strong
blueshifted emission component near the nucleus, and we find weaker
broad, high-velocity emission at greater distances.  These
high velocities, ranging from $\sim500$ km s$^{-1}$ with respect to the
systemic velocity to over twice that, cannot plausibly be due to gravitational
dynamics.  Instead, they very likely reflect the interaction of the
radio jet with the ambient medium.

As mentioned in \S \ref{pops}, some of the youngest stellar populations are
found just NE of the quasar. It was in this region that
Stockton \& Ridgway \markcite{sto91}(1991) discovered what looked like
a secondary nucleus 1\arcsec\ from the quasar nucleus, which they called 
3C\,48A.   Figure \ref{imagemontage}$g$ and $h$ show WFPC2 PC
images of 3C\,48 from the {\it HST} archive, which confirm the object found
in the ground-based imaging and show much intriguing structural detail.
While it is still possible that this may be, in fact, the distorted
nuclear regions of the companion galaxy in the final stages of merger,
it now seems more likely that it, too, is related to the interaction of the
radio jet with the dense surrounding medium, as suggested by
Chatzichristou \etal\ \markcite{cha99}(1999).

The WFPC2 PC images (Fig.~\ref{imagemontage}$g$,$h$) show an almost circular
edge brightening effect around 3C\,48A, which Kirhakos \etal\
\markcite{kir99}1999 suggest
is an image ghost artifact.  After careful inspection, we believe it to be
a real feature, especially since it perfectly surrounds the peak seen
in both the ground-based and {\it HST} imaging. The VLBI map of
3C\,48 (Wilkinson \etal\ \markcite{wil91}1991) shows diffuse radio emission
extending into this region.
Together with the spectroscopic evidence, the radio morphology
suggests that this edge-brightening effect may be the remains of a
bubble, where the radio jet had recently, but temporarily, broken through
the dense gas in the inner region.  We believe that this is not an active
bubble at present because (1) the present jet appears to be deflected along 
a more
northerly direction, and (2) the shell-like structure is not apparent in the
emission-line image, indicating that it probably comprises stars, which
may have been formed as shocks compressed the gas at the boundaries of the
bubble.

From the weakness of the core component in 3C\,48, Wilkinson \etal\
\markcite{wil91}(1991) argue that the radio jet
is truly one sided, \ie\ we are not missing an oppositely directed jet
simply because of Doppler boosting.  Our spectroscopy tends to confirm
this view:  we see widespread high-velocity gas with blueshifts relative
to the systemic velocity, but not with redshifts.

\section{Summary and Conclusions}

In summary, 

\begin{itemize}
\item We have measured redshifts from stellar absorption features in
32 regions of the host galaxy.  The average redshift in the 
central region of the host is close to $z=0.3700$, and there are relatively
small variations ($\sim50$ km s$^{-1}$) in the main body of the galaxy.
The large faint feature extending 
NW of the quasar is clearly blueshifted with respect to the main body of the 
host galaxy by as much as 300 km s$^{-1}$.   Some regions in
the SE of the galaxy are redshifted, suggesting that the system is rotating
about an axis oriented roughly NE---SW.
\item We have successfully modeled the stellar populations in the same 32 
regions of the host galaxy of 3C\,48.  Spectra from most regions can be 
modeled by an old stellar component plus an instantaneous burst population, 
presumably the stellar component present in the galaxies prior to 
interaction plus a starburst produced as a result of the interaction.
Spectra from the remaining regions are well fit by the old component alone.
\item The feature extending NW of the quasar is a tidal tail 
composed mostly of an old stellar component, with clumps of recent 
star formation, much like those observed in other merging systems.  
The estimated dynamical age of this feature (between 0.1 and 0.3 Gyr) 
is much younger than the age of the dominating stellar population 
($\sim$ 10 Gyr) and roughly equal to or older than the starburst ages we 
find anywhere on the host ($\leq$ 0.11 Gyr).  Thus the starbursts 
occurred after the tidal tail was initially launched.   This delay of
most of the star formation until close to the time of final merger
indicates that at least one of the merging galaxies probably had a massive
bulge capable of stabilizing the gas in the inner disk.
\item We find very young stellar populations in the central regions of
the host galaxy, with the youngest populations ($\sim4$ Myr) 
in a region $\sim1\arcsec$ NE of the quasar and in a region $\sim2\arcsec$
SE of the quasar. 
The youngest ages we find ($<$ 10 Myr) indicate ongoing star formation.
\item The large amounts of observed CO (Scoville \etal\ \markcite{sco93}1993; 
Wink \etal\ \markcite{win97}1997) along with the very large fractions of gas 
already used up to form stars in the starburst regions we observe indicate 
that 3C\,48 may be near the peak of its starburst activity.
\item The extended emission is distributed mostly in several discrete
regions.  Most of this gas falls in one of two distinct velocity regimes:
either (1) within $\sim200$ km s$^{-1}$ of the systemic velocity, or (2)
blueshifted with respect to the systemic velocity by $\sim500$ km s$^{-1}$.
Densities in at least the brighter clumps of extended emission are $\sim150$
$e^-$ cm$^{-3}$, indicating probable compression by shocks resulting from
the interaction.  The presence of at least one region of recent star formation
in the tail indicates that some of these clumps reach sufficient densities to
become self gravitating.
\item Extremely luminous high-velocity emission close to the quasar
nucleus and extended emission with high velocities and large velocity
widths, together with the convoluted shape of the VLBI radio jet, indicate a 
strong interaction between the radio jet and the ambient gas.  The luminosity
peak 1\arcsec\ NE of the quasar and the edge-brightening around this region
seen in the {\it HST} images may be a relic of a previous, but still quite 
recent, interaction between the radio plasma and the ambient material.
\end{itemize}

From these results, we can outline a tentative evolutionary history for the
host galaxy of 3C\,48.  Two gas-rich galaxies of comparable mass, at least one
of which has a massive bulge, interact strongly enough for tidal friction
to have a significant influence on their mutual orbit.  The orbit is prograde
with respect to the stellar disk of at least one of the galaxies, and, at 
their first close passage, a tidal tail is produced from this disk.
Little star formation occurs at this point, but the tail does include some
gas which will later form small star-forming regions, particularly along
the leading edge of the tail.
Most of star formation is delayed by a few hundreds of Myrs until the
final stages of merger, when gas-flows into the center also trigger the
quasar.  A one-sided radio jet is produced, which interacts strongly with
the dense gas that has accumulated near the center, producing a bubble of
shocked gas to the NE of the quasar.  The shocked gas at the boundary of
this bubble forms stars; shortly afterwards, the jet is deflected along
a more northerly direction.

Information on stellar populations in host galaxies and close companions
to QSOs can potentially give us a substantial lever
in attempting to sort out the nature and time scales of various phenomena 
associated with the initiation and evolution of QSO activity.
Numerical simulations suggest that the peak of starburst activity 
for mergers occurs at 
roughly the same time as a major gas inflow towards the center of the galaxy, 
so the time elapsed since the peak of the central starburst may be 
approximately coincident 
with the QSO age.  Thus, by determining starburst ages in these objects we 
may be able to place them on an age sequence, and this in turn may help
clarify the relationship between ULIGs and QSOs.

In particular, our results are suggestive of a connection between 3C\,48 and 
the ULIG population.  We have shown that 3C\,48 is likely to be near the peak 
of the starburst activity,  which would place it near the beginning of the 
age sequence mentioned above.   We have also presented evidence that 3C\,48 
is in the final stages of a merger;
ULIGs are found preferentially in the final merging phase (Surace \etal\
\markcite{sur98}1998; Mihos \markcite{mih99}1999).  Further, 3C\,48 occupies 
a place in the FIR diagram close to that of ULIGs (Neugebauer \etal\ 
\markcite{Neu86}1986; Stockton \& Ridgway \markcite{sto91}1991),
and Haas \etal\ \markcite{haa98}1998 find that the FIR emission of 3C\,48 
is unambiguously
dominated by thermal emission. Indeed the mass and dominant temperature of
dust in 3C\,48 appears to be very similar to that of ULIGs
(Klaas \etal\ \markcite{kla97}1997; Haas \etal\ \markcite{haa98}1998).
In short, if the evolutionary sequence proposed by Sanders \etal\
\markcite{san88} (1988) is correct,
3C\,48 seems to be viewed right after the optical QSO becomes visible.   

While this scenario may seem plausible, there are clearly a number of
worries.  To what extent is it legitimate to use a star-formation age
measured several kpc from the nucleus as a proxy for that in the nucleus
itself, which is the age that we might expect to be most closely correlated
with that of the QSO activity?  Given that we are not sure what 
angular-momentum
transfer mechanisms operate to bring the gas from the 100-pc-scale of the
central starburst to the sub-pc scale of the QSO accretion disk, is our
assumption that the starburst peak and the triggering of the QSO activity
are more-or-less contemporaneous valid?  

The picture also becomes more complex
when one tries to make sense of recent related observations.
Tran \etal\ \markcite{tra99}1999 concluded that QSOs are likely to be found 
only in host galaxies with a dominant stellar population older than 300 Myr.
Apparently, their reasoning is based on the observation that ULIGs have
starbursts with mean ages of $\sim300$ Myr, and the belief that the active 
nucleus
only becomes visible at a later stage.  3C\,48 is a clear counterexample to
this view.  There clearly must be a dispersion, and quite likely a rather
broad one, in the properties of ULIGs that govern the time it takes for
dust to be cleared from the inner regions.  Even more important may be the
viewing angle.  For 3C\,48, there is almost certainly a range of lines of 
sight for which the quasar is hidden and the object would be classified 
solely as an ULIG.

In a study of off-nuclear optical spectroscopy of 19 QSO host galaxies,
Kukula \etal\ \markcite{kuk97}1997 found much older dominant stellar
populations in the hosts, with ages ranging from $\sim2$ Gyr to $\geq11$ Gyr.
While there certainly do seem to be QSOs for which the host-galaxy spectrum
is dominated by old stars, our detailed study of 3C\,48 does inject a note
of caution regarding such determinations, particularly those based on a 
single slit position in the outskirts of the host galaxy.
If we were to confine our interpretation of the stellar population of 3C\,48 
to an integrated spectrum from points $\gtrsim6\arcsec$ from the quasar, 
we would have found only an old population.
However, it is clear that our selection of QSOs with FIR colors similar to
those of ULIGs has strongly biased our results; in fact, none of the
host galaxies or strongly interacting companions in this sample that 
we have yet observed (\eg\
PG\,1700+518: Canalizo \& Stockton \markcite{can97}1997; Stockton,
Canalizo \& Close \markcite{sto98}1998;
Mrk\,231, IR\,0759+651, Mrk\,1014: Canalizo \& Stockton \markcite{can00}2000)
show dominant populations as old as the {\it youngest} of those observed by
Kukula \etal\ \markcite{kuk97}1997.

On a more speculative note, there may be a correlation between the
morphology of the merging galaxies, the star-formation properties, and the
QSO properties.  If the models of Mihos \& Hernquist (1996) are at least
qualitatively correct, mergers involving disk galaxies of nearly equal mass,
but with insignificant bulges, will have a rather broad peak of star formation
in which they use up most of their available gas near the time of first
passage, leaving little for further star formation at the time of final
merger.  Similar mergers involving pairs with substantial, centrally
concentrated bulges have only a small amount of star formation prior to
their final merger, when most of the star formation occurs in quite a
sharp peak.  Thus one expects the most luminous starbursts to be in mergers
involving gas-rich galaxies with massive bulges.  Comparison of Monte-Carlo
simulations of mergers with images of a sample of ULIGS does indicate that
most of the latter are in the late stages of merger, consistent with
stability of their inner disks over the early phases of the interaction
(Mihos 1999).

While bulges may be important in determining the timing and the rate of
star formation during an interaction, they may also have some bearing on
the nature and strength of the nuclear activity.  The recent demonstration
of an apparent correlation between bulge mass and black-hole mass
(Magorrian \etal\ 1998) indicates that those mergers for which gas-flows
into the center are delayed by the presence of stabilizing bulges and that have
the highest intrinsic flow rates may also involve the most massive black
holes.  If QSO luminosities are statistically correlated with the Eddington
limits on their black-hole accretion rates, then these sorts of mergers
are likely to produce the most luminous QSOs.  While there are many
qualifications to this scenario (\eg\ current N-body simulations fall
several orders of magnitude short of being able to follow the gas to
the scale of the accretion disk; feedback from stellar winds and supernova
outflows on the infalling gas is not well treated), it does at least supply
a motivation to be alert to possible differences between QSOs for which the
star formation and nuclear activity occur while the objects are still
distinctly separated, such as PG\,1700+518 (Canalizo \& Stockton 1997;
Stockton \etal\ 1998) and those, like 3C\,48, where the activity appears
to peak close to the time of final merger.

\acknowledgments

We thank Bill Vacca, John Tonry, Dave Sanders, and Josh Barnes for 
helpful discussions, and Susan Ridgway for assisting in some of 
the observations.  We also thank Richard Hook for supplying us with his
CPLUCY routine and Matt McMaster for his detective work in confirming 
the spurious nature of the apparent object near 3C\,48 in the {\it HST}
[\ion{O}{3}] image.  We are grateful to the referee, Linda Dressel, for useful 
comments and suggestions.
This research was partially supported by NSF under grant AST95-29078. 

\newpage
\appendix
\section{Serendipitous Objects Around 3C\,48}
A number of objects in the field of 3C\,48 fell on our various slit
positions.  We measured redshifts for those objects which could also be 
identified in the {\it HST} images.  Table \ref{seren} 
lists these objects with their coordinates (J2000.0) as measured from 
the {\it HST} images, redshift, whether the redshift was determined from 
emission or absorption features, and the slit position ID (see \S\ref{obs}).

Gehren \etal\ \markcite{geh84}(1984) suggest that the object 12\arcsec\ NW 
of the quasar nucleus may be a companion galaxy to 3C\,48.   This 
object falls on the edge of our slit position E.   We see evidence for a 
faint emission line at the 2.5 sigma level, wide enough at this resolution
to be [\ion{O}{2}] $\lambda 3727$.  If this were the case, the redshift of
the object would be $z=0.378$ ($\Delta$v = $\sim 1800$ km s$^{-1}$ with 
respect to our zero velocity point in the host galaxy).   This object is
not listed on Table 3.

Only one object in Table 3 has a redshift relatively close to that of 
3C\,48.  This may indicate that 3C\,48 is in a low density cluster (Yee 
\& Green \markcite{yee87}1987), if in a cluster at all.

The background galaxy $\sim6\arcsec$ SE of 3C\,48 (object 4) stresses the 
importance of obtaining spectroscopic redshifts to discriminate between 
objects associated with QSOs and close projections.

\newpage

\begin{figure}
%\epsscale{1.0}
%\plotone{figures/fig1.ps}
\figcaption{Maps indicating slit positions, stellar radial 
velocities, and starburst ages of the host galaxy of 3C\,48.  The contour 
maps are of an image similar to that of Fig. 2$a$, and the contours are 
drawn at logarithmic steps.   In all panels N is up and E to the left, and
the tickmarks are in arcseconds.   ($a$) Region IDs:  Slit positions A, B, 
C and G
indicating the regions into which we have subdivided each slit for analysis.
($b$) Radial Velocity Map:  Shading indicates radial velocity with respect to 
$z=0.3700$ (close to the average redshift in the central regions of the host 
galaxy) measured from stellar absorption features.  ($c$) Starburst Age Map:
Shading indicates ages of instantaneous burst populations determined from
our modeling (see text for details).  Scalebar is logarithmic; three tickmarks
are drawn at 10, 50, and 100 Myr respectively.
The hatched regions indicate the predominance of an old stellar population,
\ie\ regions in which the observed spectrum is well fit by the old stellar
population model alone. \label{master}}

\figcaption{Ground-based and {\it HST} WFPC2 images of 3C\,48. 
In all cases, N is up and E to the left.  Small insets, when present, 
show lower-contrast versions.
Detail of the observations are given in \S2 of the text.  ($a$)  Image
obtained with an interference filter centered at 6120 \AA, with a FWHM of
960 \AA, covering the mostly emission-line-free rest frame region
4120--4820 \AA.  This image has been deconvolved and had the quasar profile
removed with CPLUCY (Hook 1998).  The white cross indicates the position of
the quasar.  The image prior to CPLUCY processing is shown in ($b$) and
a lower-contrast version of ($a$) is shown in ($c$)  ($d$)  Image
in the rest-frame continuum near 2500 \AA, also processed with CPLUCY.
($e$)  Ratio of the $U'$ image ($d$) to the $R'$ image ($a$)
Both images have been smoothed to the same resolution, and regions with
a S/N $<2$ in either image have been suppressed.  ($f$)  Image through
a 30 \AA\ FWHM filter centered on the redshifted [O\,III] $\lambda5007$
emission line.  ($g$---$i$)  Archival images obtained with 
{\it HST} and WFPC2 (filters as labeled; note that these images are enlarged by a factor of 2 relative to the other images).  ($g$) and ($h$) were 
obtained with PC1 and ($i$) with WFC2 and a linear-ramp filter.
\label{imagemontage}}
\end{figure}

\begin{figure}
%\epsscale{1.0}
%\plotone{figures/fig3a.ps}
%\plotone{figures/fig3b.ps}
%\plotone{figures/fig3c.ps}
\figcaption{Characteristic spectra of different regions of the host galaxy of
3C\,48. Each panel displays the observed data in rest frame (heavy line),
an underlying old stellar population model (lower light line), a young 
instantaneous starburst model (upper light line), and the $\chi^2$ fit 
of the sum of the two models to the data (dotted line).
The old model is a 10 Gyr old stellar population with an exponentially 
decreasing star formation rate with an e-folding time of 5 Gyr.   
Each panel is labeled with the region ID it represents (cf. Fig. 1).
The original data have been smoothed with Gaussian filters with $\sigma =
2$ \AA.  Emission lines seen in these spectra come from the extended-emission
region of the quasar and not from the star-forming regions. \label{typical}}
\end{figure}

\begin{figure}
%\epsscale{0.6}
%\plotone{figures/fig4.ps}
\figcaption{Estimating errors in age determination.
The observed spectrum from region
B8 (solid line) with three different solutions superposed (dotted line):  
The youngest that still looks reasonable (top), 
the best fit (middle), and the oldest that still looks
reasonable (bottom).  The ages of the young starbursts are 72, 114, and 181 
Myr respectively. In each case, the relative contribution of the old 
population was increased or decreased to give the best fit.
The difference in $\chi^2$ statistic between the middle and top or middle and 
bottom fits is $\sim$15\%.  The flux-density scale refers to the middle
spectrum; the upper and lower spectra have been displaced by arbitrary
amounts. \label{agerr}}
\end{figure}

\begin{figure}
%\epsscale{1.0}
\figcaption{Trace of the nuclear spectrum of 3C\,48, showing the double peaks
in the [O\,III] lines.  The spectrum has been reduced to the rest frame
using the redshift of the longward component, which closely matches the
broad-line redshift.  The redward side of the [O\,III] $\lambda5007$ line
is slightly affected by atmospheric B-band absorption; even more so is the
Fe\,II $\lambda5018$ emission, which appears as a small, displaced peak
to the right of the [O\,III] $\lambda5007$ line.  \label{hb_nuc_emiss}}
\end{figure}

\begin{figure}
%\epsscale{0.5}
%\plotone{figures/fig6.ps}
\figcaption{Deconvolved spectrum from slit A. This image was produced by
first subtracting off as much of the nuclear continuum as possible without
producing negative residuals, running a PLUCY deconvolution (Hook
\etal\ 1994) with no point-source components, and dividing the result by
an image smoothly varying as a function of slit coordinate and designed 
to reduce the large dynamic range in the emission component intensities.
While this procedure produces negative artifacts near strong emission 
features and gives a distorted view of the strengths of the various
emission components, it clearly shows the strong velocity gradient in the
strong anomalous narrow-line region near the nucleus. \label{oiiidecon}}
\end{figure}

\begin{figure}
%\epsscale{0.5}
%\plotone{figures/fig7.ps}
\figcaption{Slit positions superposed on the [O\,III] emission image. The inner
region is shown at lower contrast in the main panel, and the inset shows
most of the extended emission components at an intermediate contrast.
\label{emslits}}
\end{figure}

\begin{figure}
%\epsscale{1.0}
\figcaption{Two-dimensional spectral images, showing either the
[O\,III] $\lambda\lambda4959,5007$ or the [O\,II] $\lambda\lambda3726,3729$
lines.  Slit orientations are shown in Fig.~7.  Note the broad,
blue-shifted components, seen most obviously in slits B, C, and G.
\label{specimage}}
\end{figure}

\begin{figure}
%\epsscale{1.0}
\figcaption{
Overview of the properties of the emission-line gas observed in each of six
slits, and the relation of the gas velocities and stellar velocities along
the lines of sight.  The circles are centered on the velocity and position
of each emission-line component detected, and their areas are proportional
to the flux at that point.  The horizontal lines are proportional to the
Gaussian FWHM.  The scale for each of these measures is given in the
lower-left corner of each panel:  the circle corresponds to 
$10^{-16}$ erg cm$^{-2}$ s$^{-1}$ per 1-pixel row ($=0\farcs215$) in the 
spatial dimension, and the line corresponds to a FWHM of 400 km s$^{-1}$.
The points with error bars give the stellar velocities for each slit.  The
horizontal bars on these points give velocity uncertainties, but the vertical
bars give the spatial range over which the velocity is an average.  The zero
point of the spatial axis for Slit A is centered on the quasar; that for
the other slits is referenced to the point closest to the quasar.  The
asterisk (*) in the diagram for Slit F indicates a region for which emission
was observed, but multiple components and a large velocity dispersion made
measurements of the [O\,II] doublet impossible.  This region roughly 
corresponds
to that covered by Slit C for $-1\lesssim\Delta y\lesssim0$. \label{emvel}}
\end{figure}

\newpage

\begin{center}
\vspace{1in}
\begin{deluxetable}{cclccc}
\tablecaption{Journal of Spectroscopic Observations \label{journal}}
\tablehead{\colhead{} & \colhead{PA} & \colhead{Offset} & \colhead{Slit Width}
& \colhead{Dispersion} & \colhead{} \\
\colhead{Slit ID} & \colhead{(deg)} & \colhead{(arcsec)} & \colhead{(arcsec)}
& \colhead{(\AA pixel$^{-1}$)} & \colhead{UT Date} }
\startdata
A &  332.7  &  \phn\phn0.0    & 1.0  & 1.28 & 96 Oct 13  \nl
B & \phn13.5 & \phn\phn1.9 E  & 1.0  & 1.28 & 96 Oct 13  \nl
C &  300.0  &  \phn\phn2.5 N  & 1.0  & 1.28 & 96 Oct 13  \nl
D &  185.0  &  \phn\phn0.0    & 0.7  & 0.85 & 96 Oct 14  \nl
E &  100.0  &  \phn\phn3.8 N  & 1.0  & 0.85 & 96 Nov 03  \nl
F &  148.0  &  \phn\phn2.1 E  & 0.7  & 0.85 & 96 Nov 03  \nl
G &  167.0  &  \phn\phn2.2 W  & 1.0  & 1.28 & 96 Nov 04  \nl

\enddata

\end{deluxetable}
\end{center}

\begin{center}
\begin{deluxetable}{lrcccc}
%\footnotesize
\tablecaption{Stellar Populations and Kinematics \label{regions}}
\tablehead{\colhead{} & \colhead{Velocity\tablenotemark{a}}  
& \colhead{Age\tablenotemark{b}} & \colhead{} 
& \colhead{} & \colhead{\% QSO} \\
\colhead{Region ID} & \colhead{(km s$^{-1}$)}  
& \colhead{(Myr)} & \colhead{\% Mass\tablenotemark{c}} 
& \colhead{\% Light\tablenotemark{d}} 
& \colhead{Contamination\tablenotemark{d,e}} }
\startdata
A1 & $+179\pm\phn18\phn\phn$ & \phn\phd\phn50\phn  & \phn3.6 & 42  &  15  \nl
A2 & $+107\pm\phn41\phn\phn$ & \phn\phd\phn114\phn & \phn3.4 & 33  &  30  \nl
A3 & $+46\pm\phn20\phn\phn$  & \phn\phn4.8\phn     & \phn0.6 & 41  &  63  \nl
A4 & $-86\pm120\phn\phn$     & \phn\phn9.1\phn     & \phn2.9 & 63  &  61  \nl
A5 & $-148\pm102\phn\phn$    & \phn\phn9.1\phn     & \phn1.6 & 47  &  34  \nl
A6 & $-261\pm\phn50\phn\phn$ & \phn\phn8.7\phn     & \phn0.6  & 25 &  17   \nl
A7 & $-212\pm\phn78\phn\phn$ & \phn\phd\phn14\phn & \phn1.5  & 45  & \phn0 \nl
A8 & $-147\pm\phn94\phn\phn$ & \phd old         &  \phn0.0 & \phn0 & \phn0 \nl
B1 & $-2\pm\phn92\phn\phn$  &  \phn\phn8.3\phn  &  \phn1.3  & 47  &  34  \nl
B2 & $+42\pm\phn52\phn\phn$ &  \phn\phn9.1\phn  &  \phn2.3  & 56  &  46  \nl
B3 & $+20\pm\phn16\phn\phn$ &  \phn\phn4.4\phn  &  \phn1.6  & 63  &  59  \nl
B4 & $-2\pm\phn25\phn\phn$  &  \phn\phn6.3\phn  &     15.2  & 94  &  63  \nl
B5 & $+48\pm\phn18\phn\phn$ &  \phn\phn4.8\phn  &  \phn1.5  & 67  &  27  \nl
B6 & $+28\pm\phn20\phn\phn$ & \phd\phn102\phn   &     11.4  & 63  & 16   \nl
B7 & $-30\pm\phn30\phn\phn$ & \phd\phn102\phn   &  \phn7.4  & 51  & \phn6 \nl
B8 & $-33\pm\phn51\phn\phn$ & \phd\phn114\phn   &     11.0  & 60  & \phn0 \nl
C1 & $+11\pm\phn26\phn\phn$ &  \phn\phn5.5\phn  &     12.2  & 92  &  48  \nl
C2 & $+17\pm\phn33\phn\phn$ &  \phn\phn5.5\phn  &  \phn1.0  & 48  &  42  \nl
C3 & $+4\pm\phn65\phn\phn$  &  \phn\phn9.1\phn  &  \phn2.0  & 53  &  32  \nl
C4 & $-57\pm\phn21\phn\phn$ &  \phn\phn8.7\phn  &  \phn2.3  & 59  &  12  \nl
C5 & $-112\pm\phn52\phn\phn$ & \phn\phn8.7\phn  &  \phn2.0  & 54  &  \phn0 \nl
C6 & $-298\pm\phn92\phn\phn$ & \phn\phn9.1\phn  &  \phn1.4  & 44  &  \phn0 \nl
C7 & $-335\pm\phn30\phn\phn$ & \phd old  &  \phn0.0  &  \phn0  &  \phn0  \nl
C8 & $-175\pm\phn61\phn\phn$ & \phd old  &  \phn0.0  &  \phn0  &  \phn0  \nl
G1 & $+103\pm\phn74\phn\phn$ & \phn\phd\phn35\phn & \phn5.1 & 54 & \phn0 \nl
G2 & $+4\pm\phn27\phn\phn$   & \phn\phd\phn16\phn & \phn4.0 & 62 & \phn0 \nl
G3 & $-28\pm\phn14\phn\phn$  & \phn\phn9.1\phn    & \phn2.4 & 58 &  67   \nl
G4 & $+66\pm\phn44\phn\phn$  & \phd old   &  \phn0.0  &  \phn0   &  82   \nl
G5 & $-129\pm\phn60\phn\phn$ & \phd old   &  \phn0.0  &  \phn0   &  78   \nl
G6 & $-287\pm\phn30\phn\phn$ &  \phn\phn9.1\phn  & \phn1.3  & 41 & \phn0 \nl
G7 & $-325\pm\phn98\phn\phn$ & \phd old   &  \phn0.0  &  \phn0   & \phn0 \nl
G8 & $-194\pm\phn30\phn\phn$ & \phn\phd\phn33\phn & \phn7.0 & 64 & \phn0 \nl
 
\enddata
\tablenotetext{a}{Relative to a systemic redshift $z=0.3700$}
\tablenotetext{b}{Error estimates for starburst age described on text.  See
also Fig. \ref{agerr}}
\tablenotetext{c}{Assuming a Salpeter initial mass function}
\tablenotetext{d}{Light contribution from starburst and QSO contamination
measured at rest wavelength 4500 \AA }
\tablenotetext{e}{Percentage of QSO contamination in each region depends
on both distance from nucleus and surface brightness of host in that particular
region.}
\end{deluxetable}
\end{center}

\begin{center}
\begin{deluxetable}{lllccl}
\tablecaption{Serendipitous Objects \label{seren}}

\tablehead{\colhead{} & \multicolumn{2}{c}{Coords (J2000)} & 
\colhead{} & \colhead{Slit} & \colhead{} \\ 
\colhead{ID} & \colhead{$\alpha$} & \colhead{$\delta$} & 
\colhead{Redshift\tablenotemark{a}} & \colhead{Position} & 
\colhead{Comments\tablenotemark{b}}}
\startdata
1  & 1$^{\rm h}$37$^{\rm m}$39\fs04 & 33\arcdeg10\arcmin33\farcs8                  & 0.6722 e\phn & A &  interacting  \nl
2  & 1\phm{$^{\rm h}$}37\phm{$^{\rm m}$}42.33 & 33\phm{\arcdeg}09\phm{\arcmin}09.0 & 0.7408 ea & A &   disk    \nl
3  & 1\phm{$^{\rm h}$}37\phm{$^{\rm m}$}42.79 & 33\phm{\arcdeg}09\phm{\arcmin}00.0 & 0.6730 ea & A & elliptical  \nl
4  & 1\phm{$^{\rm h}$}37\phm{$^{\rm m}$}41.54 & 33\phm{\arcdeg}09\phm{\arcmin}29.0 & 0.8117 e\phn & A &   disk \nl
5   & 1\phm{$^{\rm h}$}37\phm{$^{\rm m}$}40.67 & 33\phm{\arcdeg}08\phm{\arcmin}56.8 & 2.169\phn\ e\phn & B &   BAL QSO\tablenotemark{c} \nl
6 & 1\phm{$^{\rm h}$}37\phm{$^{\rm m}$}42.15 & 33\phm{\arcdeg}09\phm{\arcmin}22.9 &    ?  & F  & continuum only \nl
7 & 1\phm{$^{\rm h}$}37\phm{$^{\rm m}$}43.96 & 33\phm{\arcdeg}08\phm{\arcmin}41.8 & 0.3560 e\phn & F &   disk \nl
8   & 1\phm{$^{\rm h}$}37\phm{$^{\rm m}$}41.67 & 33\phm{\arcdeg}09\phm{\arcmin}06.1 & 0.6430 a\phn & G & elliptical   \nl 
9    & 1\phm{$^{\rm h}$}37\phm{$^{\rm m}$}42.71 & 33\phm{\arcdeg}08\phm{\arcmin}03.1 & 0.3279 ea & G & disk \nl

\enddata
\tablenotetext{a}{$e$ and $a$ after redshift indicate whether redshift was measured from emission or absorption lines, respectively.}
\tablenotetext{b}{Morphologies are based on appearance on {\it HST} WFPC2 
images.}
\tablenotetext{c}{Canalizo, Stockton \& Roth 1998}
\end{deluxetable}
\end{center}

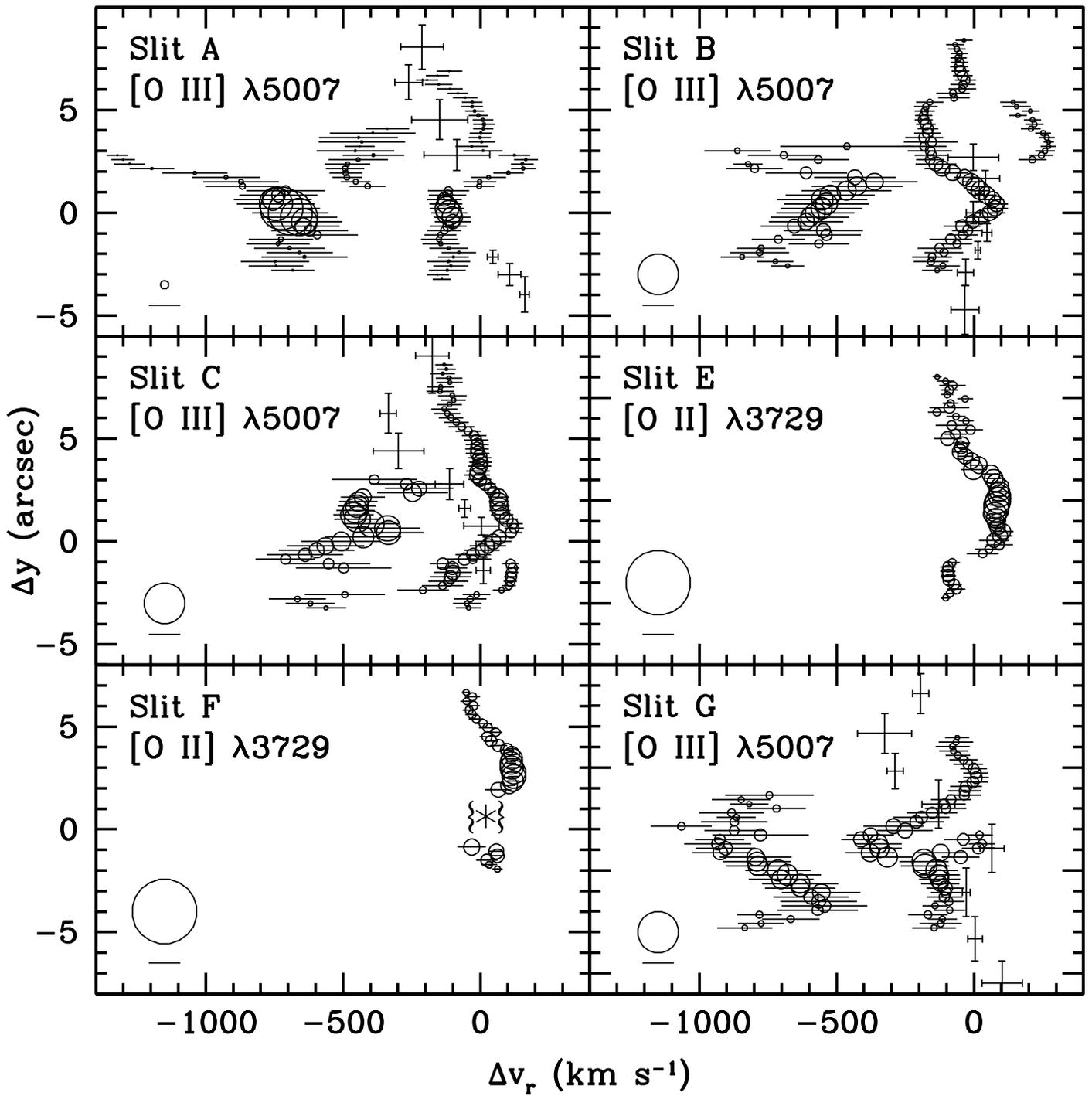
\end{document}